\documentclass[%
twocolumn,
superscriptaddress,
amsmath,amssymb,
aps,
prm,
floatfix,
]{revtex4-2}

\usepackage{graphicx}
\usepackage{dcolumn}
\usepackage{bm}
\usepackage[colorlinks=true, citecolor=blue]{hyperref}
\hypersetup{colorlinks=true,citecolor=blue,linkcolor=blue,urlcolor=blue}
\usepackage[capitalise,noabbrev]{cleveref}
\usepackage{todonotes}
\usepackage[version=4]{mhchem}
\usepackage[utf8]{inputenc}
\usepackage{multirow}
\usepackage{color}      

\newcommand{\etal}{\textit{et al.}}
\newcommand{\fh}[1]{\texorpdfstring{\ce{FeH$_{#1}$}}{FeH}}
\newcommand{\fhn}[1]{\texorpdfstring{\ce{(FeH)$_{#1}$}}{FeH}}
\newcommand{\fwu}{%
    Center for Advanced Systems Understanding, 
    D-02826 G\"orlitz, 
    Germany}
\newcommand{\hzdr}{%
    Helmholtz-Zentrum Dresden-Rossendorf, 
    D-01328 Dresden, 
    Germany}

\begin{document}

\preprint{APS/123-QED}

\title{Machine Learning-Driven Structure Prediction for Iron Hydrides}

\author{Hossein Tahmasbi}
\email{h.tahmasbi@hzdr.de}
\affiliation{\fwu}
\affiliation{\hzdr}

\author{Kushal Ramakrishna}
\affiliation{\fwu}
\affiliation{\hzdr}

\author{Mani Lokamani}
\affiliation{\hzdr}


\author{Attila Cangi} 
\email{a.cangi@hzdr.de}
\affiliation{\fwu}
\affiliation{\hzdr}

\date{\today}

\begin{abstract}
We created a computational workflow to analyze the potential energy surface (PES) of materials using machine-learned interatomic potentials in conjunction with the minima hopping algorithm. We demonstrate this method by producing a versatile machine-learned interatomic potential for iron hydride via a neural network using an iterative training process to explore its energy landscape under different pressures. To evaluate the accuracy and comprehend the intricacies of the PES, we conducted comprehensive crystal structure predictions using our neural network-based potential paired with the minima hopping approach. The predictions spanned pressures ranging from ambient to 100 GPa.
Our results reproduce the experimentally verified global minimum structures such as \textit{dhcp}, \textit{hcp}, and \textit{fcc}, corroborating previous findings. Furthermore, our in-depth exploration of the iron hydride PES at different pressures has revealed complex alterations and stacking faults in these phases, leading to the identification of several new low-enthalpy structures. This investigation has not only confirmed the presence of regions of established FeH configurations but has also highlighted the efficacy of using data-driven, extensive structure prediction methods to uncover the multifaceted PES of materials.
\end{abstract}

\maketitle


\section{\label{sec:intro}Introduction}
%
%
Iron hydride (FeH) under pressure has attracted considerable attention from geoscientists because of its relevance to the structure and properties of the Earth's inner core~\cite {anderson1986properties}. The composition of the inner core is thought to consist of iron-based alloys, but its density is slightly lower than that of pure iron~\cite{Stevenson1981,stevenson1977hydrogen,dziewonski1981preliminary}. Hydrogen is considered a critical element in the Earth's core and is thought to be the primary cause of the observed density deficit in the inner core. On the other hand, understanding how hydrogen behaves within elemental transition metals is essential for gaining insight into the properties of transition metal alloys used for hydrogen storage~\cite{Smithson2002}.
Furthermore, among the transition metal hydrides, FeH exhibits intriguing properties such as phase transformations, magnetism~\cite{Bouldi2018,tsumuraya2012first}, and the formation of vacancies~\cite{Hiroi2005}. 
 
%
Stable solid FeH has not been observed under ambient conditions. However, both theoretical and experimental studies have shown that the \textit{dhcp} phase of FeH remains stable at low pressures and undergoes phase transitions to the \textit{hcp} and \textit{fcc} phases as the pressure increases to intermediate and high levels. 
While previous \textit{ab-initio} studies suggested the stability of the \textit{hcp} phase~\cite{Skorodumova2004} instead of the experimentally observed \textit{dhcp} phase at low pressures~\cite{badding1991high,hirao2004compression,mao2004nuclear}, Isaev~\etal~\cite{isaev2007dynamical} provided an insightful solution to this discrepancy by considering the inclusion of free energy contributions in the total energy.

In recent years, there has been considerable interest, from both theoretical and experimental perspectives, in exploring the potential energy landscape and studying the equation of state (EoS) of FeH. 
Theoretically, extensive investigations have been carried out using the method of evolutionary crystal structure prediction.
Bazhanova~\etal~\cite{bazhanova2012fe} systematically explored the energy landscape of Fe$_\textnormal{x}$H$_\textnormal{y}$ (where $\textnormal{x}$, $\textnormal{y}$ = 1--4) at high pressures (300--400 GPa) corresponding to the conditions in Earth's inner core.
Following a similar approach, Kvashnin~\etal~\cite{Kvashnin2018} used the same methodology to map out the potential energy surface (PES) of FeH systems at pressures including 0, 50, and 150 GPa, focusing on superhydrides.
Sagatova~\etal~\cite{sagatova2020phase} predicted structures of FeH using a random search algorithm in the pressure range of 100--400 GPa. In addition, much effort has been devoted to predicting the crystal structures of superhydrides \fh{n} (where $n\ge3$), some of which possess intriguing properties such as superconductivity~\cite{zarifi2018crystal,Kvashnin2018,Li2017}.
Recently, Yang~\etal~\cite{Yang2022} systematically investigated the PES of FeH using evolutionary crystal structure prediction methods and density functional theory (DFT) calculations at 300 GPa and for various Fe/H ratios. These comprehensive theoretical studies shed light on the existence of a wide range of polymorphs, including both stoichiometric and non-stoichiometric FeH. 

Significant progress has already been made in synthesizing the stoichiometric FeH system by using \textit{in situ} X-ray diffraction techniques. 
These studies have demonstrated the stabilization of FeH in three distinct phases: \textit{dhcp}, \textit{hcp}, and \textit{fcc}~\cite{badding1991high,Fukai1993,Yamakata1992,Kato2020}. 
Among these phases, the \textit{dhcp} phase has been confirmed to be stable at room temperature up to 80 GPa~\cite{badding1991high,hirao2004compression}. 
In addition, investigations have revealed that the \textit{fcc} phase remains stable in the pressure range from 19 GPa to at least 68 GPa at room temperature~\cite{Narygina2011}.
In recent years, several studies have been conducted to synthesize higher hydrides of iron under pressure. P\'epin~\etal~\cite{pepin2014new} successfully identified two novel phases, FeH$_2$ and FeH$_3$, which are stable at pressures of 67 GPa and 86 GPa, respectively. 
They also performed a comprehensive reassessment of the stability of the \textit{dhcp} phase up to a pressure of 136 GPa.

%
%
From both theoretical and experimental perspectives, previous studies have mostly focused on the properties of the \textit{dhcp}, \textit{hcp}, and \textit{fcc} phases of FeH at different pressures and temperatures, such as the electrical conductivity~\cite{he2022superionic,ohta2019electrical}, sound velocity and lattice parameters~\cite{shibazaki2012sound}, thermodynamics~\cite{pepin2014new} and transport properties~\cite{umemoto2015liquid}.
Furthermore, the EoS of stoichiometric phases of FeH has been investigated thoroughly in prior work~\cite{badding1991high,Elssser1998,pepin2014new,Gomi2018,Tagawa2022,Piet2023}.

%

In this study, we present a comprehensive theoretical investigation of the PES of FeH over a wide pressure range of up to 100 GPa. To achieve this, we employ a highly transferable machine-learned interatomic potential, developed using a hierarchical approach that integrates neural networks (NNs) and the minima hopping (MH) method. By combining these advanced techniques, we demonstrate an automated and systematic methodology for training and validating transferable machine-learned interatomic potentials using global optimization methods. 
To the best of our knowledge, the PES of FeH has not been systematically studied over a wide pressure range. Therefore, we have investigated the stoichiometric FeH system in a case study.

Due to the efficiency and accuracy of the NN interatomic potential, we can perform large-scale structure predictions using the MH method, enabling us to predict both stable and metastable FeHs in a wide pressure range. 
To further refine our predictions and assess the dynamical stability of selected structures, we complement our computational framework with DFT calculations. This comprehensive approach provides valuable insights into the PES of FeH and allows us to discover several stable and meta-stable phases that have not been previously discussed. 
Furthermore, we compare our DFT results with literature data on various properties, including the bulk modulus, lattice parameters, phonon dispersion, and electrical conductivity.

\section{\label{sec:method}Computational Methods}
\subsection{Density Functional Theory}
To conduct our comprehensive DFT calculations, which involve tasks such as data set preparation for training the NN potentials, refining the MH results through local geometry optimization, and performing phonon calculations, we employ the projector augmented wave formalism as implemented in the \texttt{VASP} code~\cite{Kresse1993,Kresse1999,Kresse199615,Kresse1996}. 
We use the PBE exchange-correlation functional~\cite{Perdew1996}, taking into account the eight valence electrons of Fe atoms $3d^74s^1$ and one electron of H atoms $1s^1$. 
To ensure convergence, we utilize a plane-wave cutoff energy of $800$ eV, a force threshold $10^{-4}$ eV/\AA, and \emph{k}-point mesh with a spacing of 0.2 \AA$^{-1}$. 
A \emph{k}-point mesh with a density of 0.02 Bohr$^{-1}$ is used for the phonon calculations. 
In addition, we performed (collinear) spin-polarized calculations, setting the magnetic moments to four for iron atoms and zero for hydrogen atoms.

For the evaluation of transport properties, the real-time formulation of time-dependent density functional theory (RT-TDDFT)~\cite{Yabana1996} is utilized to compute the electrical conductivity (electronic component) based on the microscopic form of Ohm's law~\cite{Ramakrishna2023,Ramakrishna2023iop}. 
The calculations are performed using an all-electron full-potential linearized augmented plane wave (FP-LAPW) method~\cite{singh2006planewaves} as implemented in the in-house version of the \texttt{ELk} code~\cite{elk}. 
The calculations were performed using a PBE exchange-correlation functional~\cite{Perdew1996}. 
The DC conductivity is extracted by fitting a Drude form to the frequency-dependent conductivity at low energies.

\subsection{Neural Network Potential}
\subsubsection{Training}
\begin{figure}[!htbp]
\centering
\includegraphics[width=1.0\columnwidth]{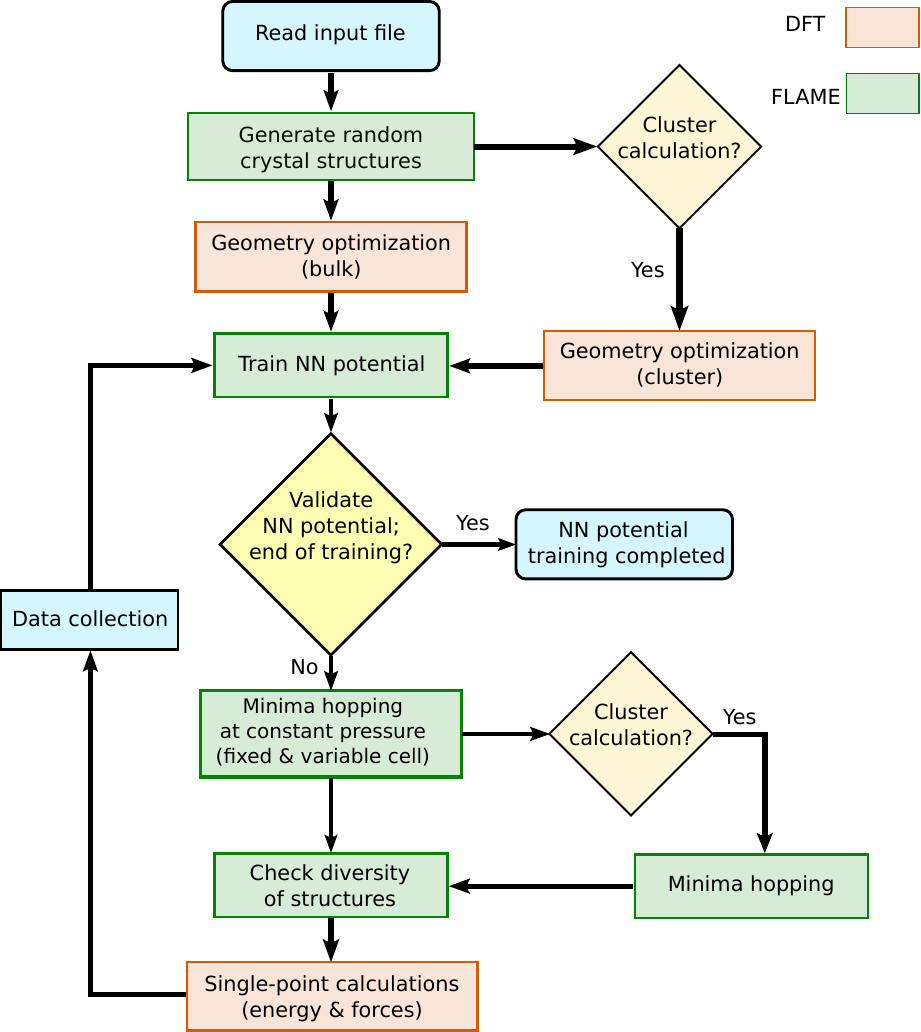}
    \caption{Workflow diagram of the automated approach for developing neural network interatomic potentials with \texttt{PyFLAME}} 
\label{fig:pyflame-workflow}
\end{figure}

To systematically explore the PES of stoichiometric FeH over a pressure range of 0--100 GPa, we train a highly transferable interatomic potential based on NNs~\cite{Behler2007} to accurately model atomic interactions. Achieving this requires training the potential using a hierarchical approach with very diverse reference datasets. 
To accomplish this, we employ the \texttt{PyFLAME} code~\cite{Mirhosseini2021}, which automatically integrates the \texttt{FLAME} code~\cite{Amsler2020} for data generation and potential construction iteratively. As \cref{fig:pyflame-workflow} shows, this iterative process involves training the NN potential and conducting crystal structure predictions to generate data to enhance the quality of the potential.
In this work, our training process using \texttt{PyFLAME} involves six sequential steps. 
In the first step, we generate random crystal structures with diverse symmetries using features implemented in \texttt{FLAME}. Subsequently, these structures are expanded, compressed, and transformed into cluster structures, forming the initial dataset for training the initial NN potential. The energies and forces of the reference datasets are calculated using the DFT.
By combining this potential and the MH method~\cite{Amsler2010}, we perform a global optimization at P$=0$ GPa to explore the high-dimensional PES of \fhn{n} clusters, where $n = $ 13, 23, 25, and 32, as well as crystal structures with $n$ values of 4, 6, 10, 13, 15, 23, 25, and 32.
After performing single-point DFT calculations, we identify and eliminate potential duplicate and high-energy structures. The refined data structures are then incorporated into the initial dataset for training a NN potential in the next step. This iterative process is repeated for five additional iterations, with the distinction that for each subsequent step, we increment the pressure for the global optimization. Specifically, we apply pressures of 10, 20, 40, 50, and 80 GPa to carry out MH runs. It is worth emphasizing that in each step, 1000 MH runs are executed for both clusters and crystals, with MH steps of 30 and 50, respectively.

In the final step, we successfully develop a highly transferable NN potential by training it on an extensive and diverse dataset. This dataset includes 33,338 clusters and crystalline structures of varying sizes and symmetries across a broad pressure range.
To ensure rigorous validation, a random selection of 20\% of the data is reserved for the validation dataset.
For the initial three steps, we utilize the NN architecture 70-14-14-1, which consists of 70 symmetry functions as input layers. These input layers comprise 16 radial functions and 54 angular functions~\cite{Behler2011}. The architecture includes two hidden layers, each containing 14 nodes, followed by one output layer. In this configuration, the output layer represents the energy of the specific atom. The input layer receives symmetry functions as descriptors of the local environment with a cutoff radius of 11.6 Bohr. 
For the last three steps, we used the architecture 70-20-20-1. 
After the final training step, we achieved remarkable results with root mean square errors (RMSEs) of less than 30 meV/atom for energy and 0.308 eV/\AA~for atomic forces. 
The specific parameters and weights of the trained potential for FeH are available in the \texttt{PyFLAME} code~\cite{pyflame}.
%

\subsubsection{Validation}
To assess the quality and reliability of the NN potential, we employ it to analyze the PES of FeH crystal structures of varying sizes. Using the MH method, we explore these structures across a pressure range from 0 to 100 GPa, starting from randomly generated structures. 
Our findings demonstrate that the NN potential accurately reproduces the results reported in the literature by identifying the local minima on the PES at different pressures. 
Nevertheless, owing to the high RMSE of energy, due to the high diversity of the dataset, our NN potential is unable to accurately determine the energetic ordering of local minima at certain pressures.

\subsection{Structure Search}
In a systematic approach, we extensively investigate the energy landscape of FeH bulk phases utilizing the MH method. 
By using a NN interatomic potential within the MH method, we achieve efficient global optimization in modeling the PES of FeH for larger length scales and across a wider range of pressures than feasible with conventional methods. 
This approach is particularly advantageous because it is computationally efficient. This allows us to discover new structures that may only be apparent at larger length scales, which would be very expensive to simulate using traditional methods. This is due to the fact that the number of local minima increases exponentially with the number of atoms in the simulation unit cell.

In the following, we explore stable stoichiometric phases of \fhn{n} in the pressure range from 0 to 100 GPa, with steps of 10 GPa. To conduct our investigations, we employ simulation cells containing 8, 10, 12, 14, 16, and 18 formula units (f.u.), representing supercells ranging from 16 to 36 atoms.
For each simulation cell size and pressure, we perform 32 MH runs with random starting configurations to thoroughly search the PES.
Our MH runs reveal very dense spectra of structures, i.e., more than a thousand structures are found at each pressure and unit cell size with relative total energies less than $200$ meV/atom compared to the global minimum structure.
After this initial sweep, we start the precise selection process. We go through the set of simulation cell sizes and pressure and remove any structure with a relative total energy exceeding $50$ meV/atom. Subsequently, duplicate structures are identified and removed by comparing their energies and space groups.
Specifically, structures with the same space group are considered distinct if the difference in their total energy exceeds $10^{-4}$ Ha. This careful elimination process ensures the integrity and uniqueness of the selected structures.
After this initial selection, the chosen structures undergo further refinement such as geometry optimization and dynamical stability calculation at the DFT level. 
Consequently, for the subsequent analysis, we consider only those structures that have a relative enthalpy of less than $30$ meV/atom compared to the global minimum structure at each pressure.

\section{\label{sec:res}Results and Discussion}
\subsection{Energetic Stability}
Through our comprehensive investigation of the PES of FeH, we discover very dense spectra of low-energy (meta)stable solid phases. Some of these structures have been reported either in the literature~\cite{Elssser1998,Skorodumova2004,isaev2007dynamical} or in material databases~\cite{Jain2013,Saal2013}. We also found several yet unknown structures across the pressure range we considered.

\begin{figure}[htp] 
\centering       
\includegraphics[width=1.0\columnwidth]{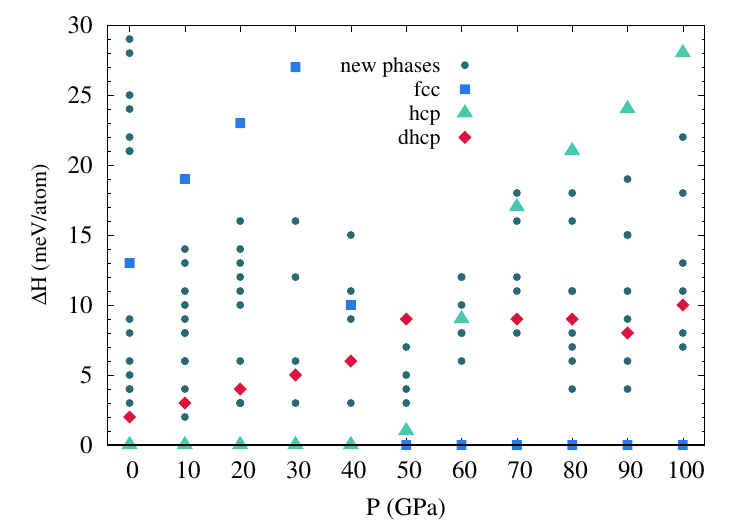}
    \caption{Calculated enthalpies as a function of pressure for the Fe-H structures without considering the free-energy corrections. The relative enthalpy shows the distance from the convex hull as a function of P. The green-filled circles represent all the new low-enthalpy stacking fault phases of FeH at each pressure found in this work.
} 
\label{fig:phase_diagram_conv}
\end{figure}

In \cref{fig:phase_diagram_conv}, we compare the relative enthalpies as a function of pressure for all the structures uncovered throughout this study. The same plot for the relative energies is provided in Figure S1 of the Supplemental Material~\cite{sup_mat}.
At pressures up to 50 GPa, there exist multiple low-enthalpy structures that lie energetically between \textit{hcp} and \textit{dhcp}. Also, at pressures above 50 GPa, one can see low-enthalpy structures between \textit{fcc} and \textit{dhcp}. It is worth mentioning that, even though our MH runs could successfully find the \textit{dhcp} on the PES of FeH at 60 GPa, our DFT calculations for this structure at that pressure did not reach convergence.
Moreover, at pressures below 40 GPa, our findings establish that the \textit{hcp} structure stands as the global energy minimum (with an energy advantage of less than $5$ meV over \textit{dhcp}), aligning with prior theoretical investigations~\cite{Skorodumova2004}. 
Nonetheless, this contradicts the experimental results~\cite{badding1991high,hirao2004compression,mao2004nuclear}, which favor the \textit{dhcp} phase as the most stable structure under low-pressure conditions.

\subsection{Structural Properties}
In \cref{tab:FeH-p20}, we present our findings for a pressure of 20 GPa, which particularly includes many unknown low-energy structures belonging to both trigonal and hexagonal crystal systems. Note that the structural data for each pressure point is provided in separate tables in the Supplemental Material~\cite{sup_mat}.   
The identified low-energy structures are labeled from S01 to S14, and the corresponding space groups, phase names, unit cell sizes, and relative enthalpies and energies are listed for each structure.

\begin{table}[tbp]
\centering
    \caption{Minima structures of FeH at P$=20$ GPa. Columns 1–3 contain the label, the space group, and the phase name, respectively. Column 4 contains the number of atoms in the unit cell of FeH. Columns 5 and 6 contain the relative enthalpy and energy with 
    respect to the global minimum structure S01 in (eV/atom).}
    \label{tab:FeH-p20}
    \begin{ruledtabular}
    \begin{tabular}{cccccc}
    Label & Space group & Phase & N & $\Delta$H & $\Delta$E \\ 
    \colrule
    S01  & $P6_3/mmc$ (194)      & \textit{hcp}  &16  & 0.000  & 0.000\\  
    S02  & $P\bar6m2$ (187)      & N8            &24  & 0.003  & 0.003\\  
    S03  & $R\bar3m$ (166)       & N1            &20  & 0.003  & 0.004\\  
    S04  & $P\bar6m2$ (187)      &               &16  & 0.003  & 0.001\\  
    S05  & $P6_3/mmc$ (194)      &               &16  & 0.003  & 0.002\\  
    S06  & $P6_3/mmc$ (194)      & \textit{dhcp} &16  & 0.004  & 0.003\\  
    S07  & $R\bar3m$ (166)       & N2            &24  & 0.006  & 0.005\\  
    S08  & $R\bar3m$ (166)       &               &20  & 0.010  & 0.005\\  
    S09  & $R\bar3m$ (166)       & N4            &24  & 0.011  & 0.009\\  
    S10  & $R\bar3m$ (166)       &               &16  & 0.012  & 0.009\\  
    S11  & $P6_3/mmc$ (194)      & N7            &24  & 0.013  & 0.008\\  
    S12  & $P\bar3m1$ (164)      &               &20  & 0.014  & 0.011\\  
    S13  & $R\bar3m$ (166)       &               &16  & 0.016  & 0.010\\  
    S14  & $Fm\bar3m$ (225)      & \textit{fcc}  &32  & 0.023  & 0.016\\  
    \end{tabular}
    \end{ruledtabular}
\end{table}

\begin{figure}[!htbp]
\centering       
\includegraphics[width=1.0\columnwidth]{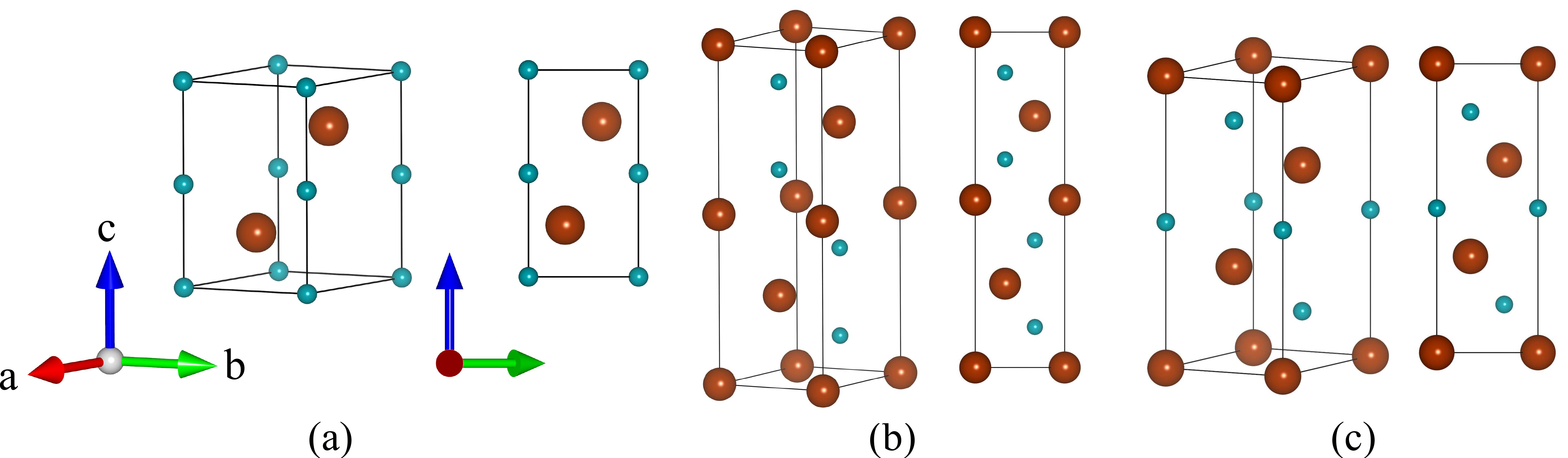}
    \caption{Structures of stoichiometric FeH reported in the literature: primitive unit cells of (a) \textit{hcp} and (b) \textit{dhcp} phases, and (c) unit cell of \textit{fcc} phase. 
    Dark red (large) and blue (small) spheres denote Fe and H atoms, respectively.}
\label{fig:dhf-phases}
\end{figure}

In addition to the well-known configurations \textit{dhcp}~\cite{Ppin2014}, \textit{hcp}, and \textit{fcc}~\cite{Kato2020} illustrated in \cref{fig:dhf-phases}, our MH runs have unveiled several novel low-energy polymorphs for large simulation cells containing up to 36 atoms.
These polymorphs either represent modifications of the known structures or are stacking faults of these. 
These findings suggest the feasibility of encountering regions in real samples that exhibit coexisting stacking sequences of \textit{fcc}, \textit{dhcp}, and \textit{hcp} under extreme external pressure and validate earlier speculations regarding the presence of coexisting regions containing known FeH structures, as discussed in prior research by Els\"asser~\cite{Elssser1998}~\etal~ 
Such confirmation underscores the benefits of uncovering the PES through large-scale global optimization techniques employing MH and NN potentials.

These new structures consist of FeH$_6$ octahedra and trigonal prism motifs, resembling the motifs found in certain binary materials~\cite{Tahmasbi2021}. In comparison to the \textit{fcc} phase, the connectivity of the octahedra in these phases undergoes a transition from edge-sharing to face-sharing due to their distinct orientations being flipped. These motifs are stacked on top of each other in various sequences and directions \textit{AB, ABC, ABCD,} \ldots.

\begin{figure}[!htbp] 
\centering       
\includegraphics[width=1.0\columnwidth]{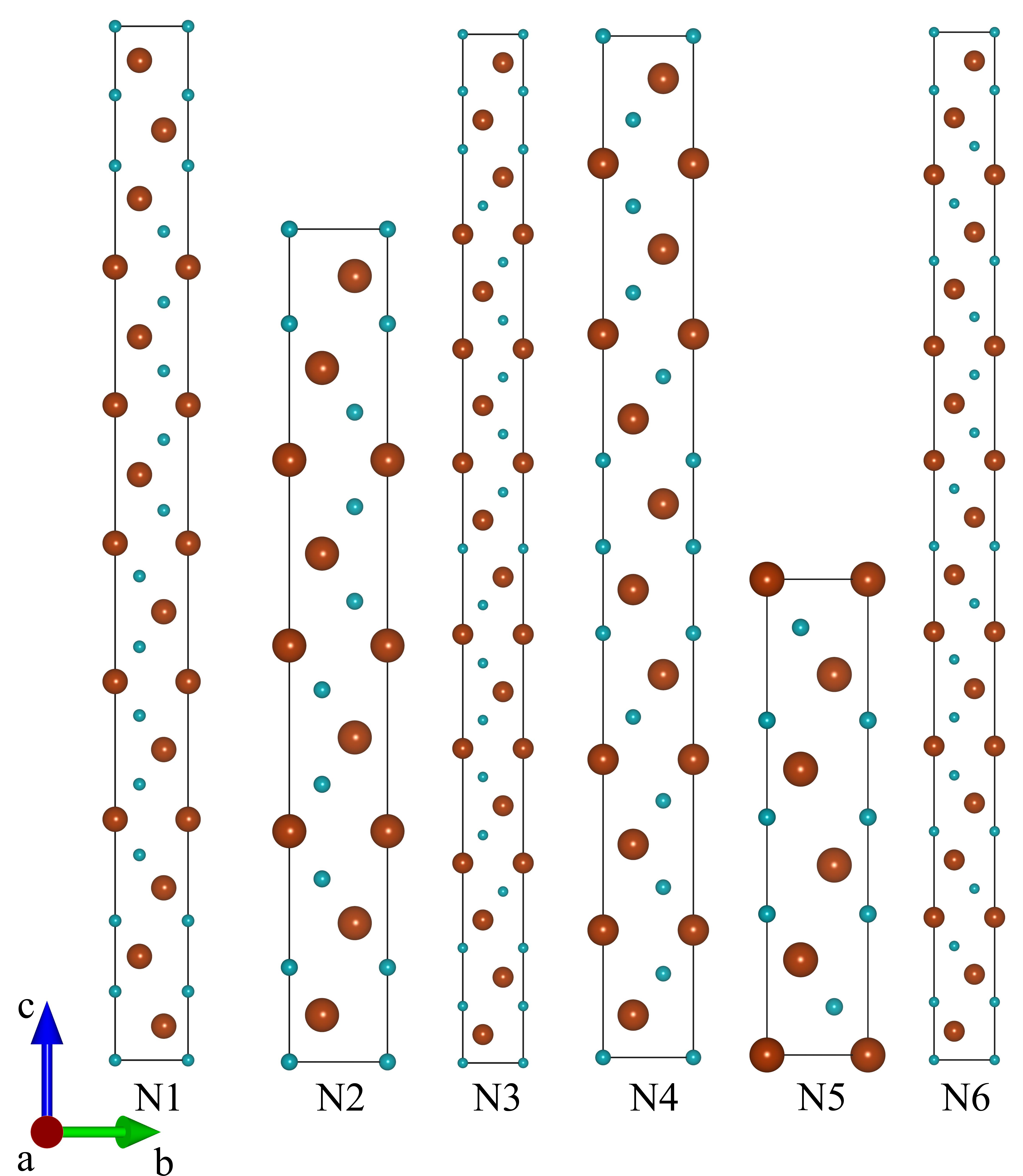}
    \caption{Low-enthalpy phases N1--N6 are new modifications or stacking faults of FeH 
    that crystallizes in the trigonal crystal system with space groups 166 ($R\bar3m$) 
    (other than N5, with space group 164 ($P\bar3m1$)).
    Some of these structures are found across all pressures 
    (see Supplemental Material~\cite{sup_mat}).}
\label{fig:new_phases2}
\end{figure}

As depicted in \cref{fig:new_phases2}, we present six novel low-enthalpy phases, denoted as N1 to N6, for FeH. These structures crystallize in the trigonal crystal system and share the same space group, 166 (except for N5, which has a space group of 164), consistently across all pressure ranges. As one can see, these phases crystallize with stacking sequences of \textit{fcc} and \textit{dhcp} and/or \textit{hcp} phases. 
For instance, N1 consists only of \textit{fcc} and \textit{hcp} phases, whereas the primitive unit cell of N2 comprises all three phases.
Notably, within the low-pressure range of 0 to 50 GPa, the N1, N2, and N5 phases exhibit remarkably low-enthalpy levels, placing them in close competition with the \textit{hcp} and \textit{dhcp} phases, with an energy difference as small as $1$ meV according to our calculations. Moreover, at certain pressures such as 10 and 80 GPa, we have also identified distinct low-enthalpy trigonal phases, each characterized by unique space groups, including 156, 160, 164, and 166.

\begin{figure*} 
\centering       
\includegraphics[width=2.0\columnwidth]{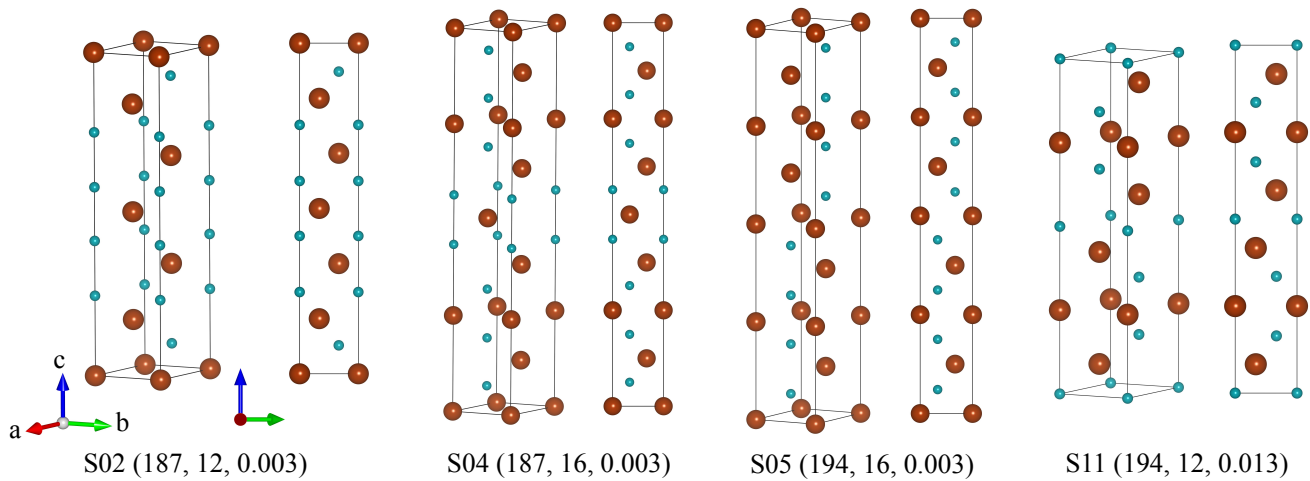}
    \caption{Some of the newly low-enthalpy modifications of FeH that crystallize in the hexagonal crystal system. They are characterized by their space groups, the number of atoms in their primitive cell, and relative enthalpies in (eV/atom) with respect to 
    \textit{hcp} at P$=20$ GPa.}
\label{fig:new_phases3}
\end{figure*}

Furthermore, as illustrated in \cref{fig:new_phases3}, some of the stacking fault structures adopt a hexagonal crystal system similar to \textit{hcp} and \textit{dhcp}, characterized by space groups 187 and 194. Specifically, the S05 and S11 (designated as N7 phase) structures represent stacking faults of \textit{hcp} and \textit{fcc} phases, respectively. 
On the other hand, S02 (the N8 phase) and S04 are composed of a combination of these two phases with additional motifs between them.
At a pressure of 20 GPa, structures S02, S04, and S05 exhibit an energetical degeneracy with the same relative enthalpies, namely a difference of only $3$ meV compared to \textit{hcp}. 
Additionally, the N7 phase, while exhibiting a relative enthalpy of 13 meV compared to the global minimum structure at 20 GPa, emerges as the second lowest energy phase at pressures of 80 and 90 GPa, with a relative enthalpy of $4$ meV in relation to the \textit{fcc} phase. 
However, this outcome is unsurprising given its structural nature as a stacking fault of the \textit{fcc} phase.

\subsection{Phonon Dispersion and Thermal Properties}
\begin{figure*} 
\centering       
\includegraphics[width=2.0\columnwidth]{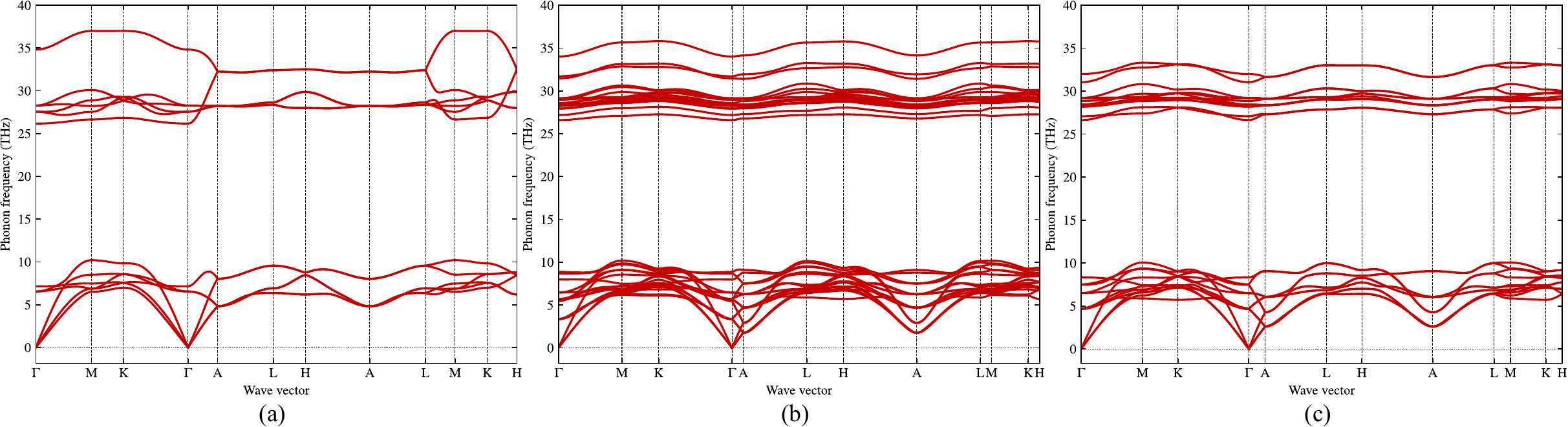}
    \caption{A comparison of the PBE phonon band structures for (a) \textit{hcp}, (b) N8, and (c) \textit{dhcp} phases at P$=20$ GPa. The LO-TO splitting was neglected.}
\label{fig:phonons3}
\end{figure*}

\begin{figure}  
\centering       
\includegraphics[width=1.0\columnwidth]{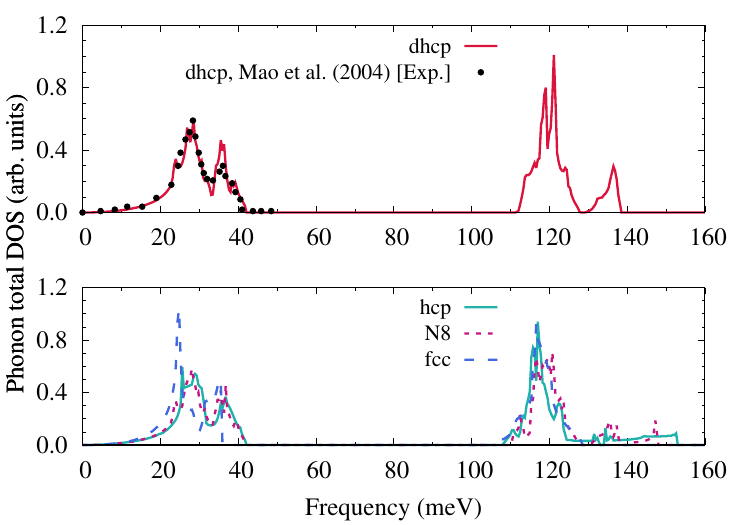}
    \caption{Comparison of the phonon DOS of the four phases of FeH at $20$ GPa with the experimental data from Ref.~\cite{Mao2004} at $22$ GPa.}
\label{fig:phonon_dos}
\end{figure}

To assess the dynamical stability and compute thermal properties, including the vibrational contribution to the free energy, we performed phonon dispersion calculations for select lowest-energy structures (refer to the Supplemental Material~\cite{sup_mat}). 
The frozen phonon approach implemented in the PHONOPY package~\cite{Togo2015} was utilized for this analysis. 
Supercells with dimensions of $2\times2\times2$, $3\times3\times3$, and $4\times4\times4$ were employed, varying based on the sizes of the primitive cells for each structure. Note that we do not account for the Born effective charge in our phonon calculations, resulting in the absence of longitudinal optical and transverse optical (LO-TO) splitting in the phonon dispersion curves. In ~\cref{fig:phonons3} we compare the phonon dispersion for three distinct structures, \textit{hcp}, N8, and \textit{dhcp}, at a pressure of P$=20$ GPa.
Additionally, we compare the phonon total density of states (DOS) for the three well-known FeH phases and N8 at 20 GPa with the corresponding experimental data at 22 GPa~\cite{Mao2004}, as illustrated in \cref{fig:phonon_dos}. It confirms the similarity of the band structures of \textit{hcp}, N8, and \textit{dhcp} phases up to 10 THz ($\approx$40 meV).

We have computed thermal properties at constant pressure using the quasi-harmonic approximation (QHA), employing the \texttt{phonopy-qha} script within PHONOPY~\cite{Togo2010}. 
In the effort to minimize the Gibbs free energy concerning volume using the integral form of the Birch-Murnaghan EoS~\cite{Birch1947,murnaghan1944compressibility}, we utilized eight volume points. 
In the QHA calculations, our focus was only on four distinct FeH phases at P$=10$ GPa, namely \textit{hcp}, \textit{dhcp}, \textit{fcc}, and N1. To this end, supercells containing $3\times3\times3$, $2\times2\times2$, $4\times4\times4$, and $2\times2\times2$ primitive unit cells are used, respectively. 
Notably, our findings reveal that, upon incorporating the free energy corrections accounting for the vibrational term $F_{\text{phonon}}(T,V)$, the energy differences and subsequently the energetic ordering of the structures remain almost unaffected, contrasting with the prior results reported by Isaev~\etal~\cite{isaev2007dynamical}. This divergence could potentially be attributed to the utilization of different DFT codes and pseudo-potentials in our study.

\subsection{Elastic and Thermodynamic Properties}
The isothermal bulk modulus results at a pressure of $10$ GPa and temperature of $300$ K for the aforementioned structures are listed in~\cref{tab:bulk_moduli} in comparison with the previous theoretical and experimental results. 
At room temperature, the volume, bulk modulus, and pressure derivative exhibit striking similarities between \textit{hcp} and N1.
The variation in the bulk modulus with temperature is illustrated in Figure S2. As the temperature rises, there is a noticeable decrease in the bulk modulus for these phases. While the distinctions between the bulk moduli of \textit{hcp}, \textit{dhcp}, and N1 are minimal at lower temperatures, these differences become more pronounced with increasing temperature. Specifically, the bulk modulus for \textit{dhcp} becomes significantly smaller than that of the other phases. 
Additionally, it is noteworthy that the bulk modulus for \textit{fcc}, which initially differs more substantially from the other phases at lower temperatures, undergoes a modest decline with increasing temperature. 
In particular, at temperatures around 500 K, 600 K, and 700 K, it converges to the same values as \textit{dhcp}, \textit{hcp}, and N1, respectively.

\begin{table}[tbp]
\centering
    \caption{Comparison of volume and bulk modulus for four FeH phases at P$=10$ GPa and $T=300$ K with corresponding theoretical (nonmagnetic) and experimental data from the literature.}
    \label{tab:bulk_moduli}
    \begin{ruledtabular}
    \begin{tabular}{ccccc}
        Phase        & V$_0$ (\AA$^3$/f.u.)   & K$_0$ (GPa) & K$_0^{'}$ & Reference    \\ 
    \colrule
        \textit{hcp} & 12.857  & 207.315      & 5.207    &This work                    \\
                     & 12.73   & 246          & 4.3      &Theo.\cite{Elssser1998}      \\  
    \hline
            N1       & 12.864  & 207.887      & 5.295    &This work                    \\  
    \hline
        \textit{dhcp}& 12.896          & 203.362       & 5.341         & This work                      \\
                     & 12.69           & 248           & 4.3           & Theo.\cite{Elssser1998}        \\
                     & 13.9$\pm$0.125  & 121$\pm$19    & 5.31$\pm$0.9  & Exp.\cite{badding1991high}     \\  
                     & 13.875$\pm$0.5  & 147$\pm$6     & 4             & Exp.\cite{hirao2004compression}\\  
                     & 12.587          & 227.2         & 4.8           &Theo.\cite{pepin2014new}        \\
                     & 13.901          & 131.1$\pm$3   & 4.83          &Exp.\cite{pepin2014new}         \\
    \hline
        \textit{fcc} & 13.026  & 195.541      & 3.957    &This work                    \\ 
                     & 12.69   & 248          & 4.3      &Theo.\cite{Elssser1998}      \\  
    \end{tabular}
    \end{ruledtabular}
\end{table}

\begin{figure*}
\centering       
\includegraphics[width=1.8\columnwidth]{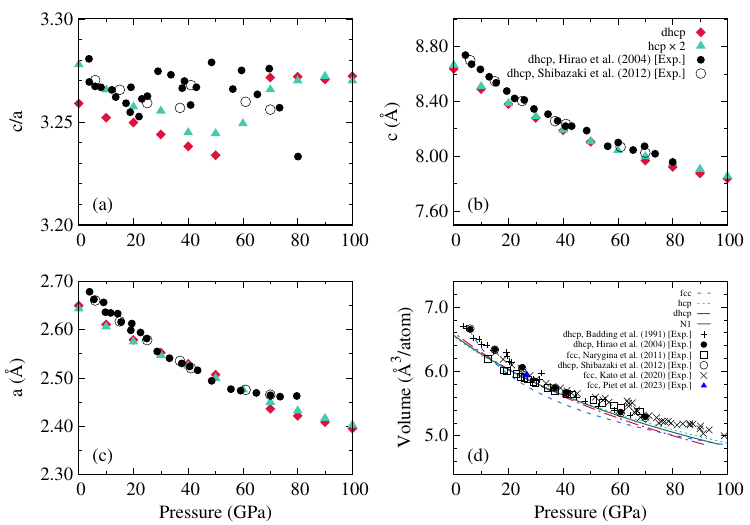}
    \caption{(a) The axial ratio ($c/a$) and (b,c) lattice parameters of \textit{dhcp} and \textit{hcp} FeH as a function of pressure in comparison with the experimental results~\cite{hirao2004compression,shibazaki2012sound}. (d) Equation of state for various phases of FeH. Experimental data for \textit{dhcp} and \textit{fcc} phases stems from Refs.~\cite{shibazaki2012sound,hirao2004compression,badding1991high,Kato2020,Narygina2011,Piet2023}.}
\label{fig:ca_ratio}
\end{figure*}

The axial ratio ($c/a$) and lattice parameters of \textit{dhcp} and \textit{hcp} FeH are plotted in \cref{fig:ca_ratio}(a-c) as a function of pressure. 
Our results agree with the experimental results~\cite{hirao2004compression,shibazaki2012sound}.
However, the $c/a$ ratio for both structures exhibits a consistent decrease with increasing pressure, extending up to 50 GPa, after which it shows a slight increase up to 100 GPa. 
Conversely, the experimental data reveals a notable discontinuity at 30 GPa, marked by a small increase in this ratio.

\cref{fig:ca_ratio}(d) shows the EoS for various FeH phases at ambient temperature. It is obtained by fitting the total energies and volume calculated using \texttt{VASP} to the third-order Birch-Murnaghan EoS~\cite{Birch1947,murnaghan1944compressibility}. 
While the three phases (\textit{dhcp}, \textit{hcp}, and N1) demonstrate identical results at low pressures, a deviation in values becomes evident at pressures exceeding 50 GPa. 
Notably, the curve for N1 consistently falls between the other two structures.
In comparison, the experimental EoS for the \textit{dhcp} and \textit{fcc} phases are additional shown~\cite{shibazaki2012sound,hirao2004compression,badding1991high,Kato2020,Narygina2011,Piet2023}. 

\subsection{Electronic Transport Properties}
Beyond EoS data, the geophysical processes occurring in the planetary interior are closely linked to the transport characteristics of iron, including its electrical and thermal conductivity. 
Most notably, the dynamo activity that produces the Earth's magnetic field is driven by the heat transfer between the planetary core and mantle~\cite{stacey2007revised,labrosse2003thermal}.
In the following, we compute the electrical conductivities of different FeH phases and compare them with the electrical conductivity of pure iron. This comparison illustrates the strong sensitivity of the electronic transport properties to structural changes.  
Note that data on the electronic transport properties of FeH under high-pressure conditions is sparse due to the difficulties in accurate measurements combined with sample synthesis commonly performed in diamond-anvil cell (DAC)~\cite{badding1991high}. Ohta~\etal~\cite{ohta2019electrical} measured the electrical resistivity (reciprocal of electrical conductivity) of the \textit{fcc} phase FeH$_{\textnormal x}$ at high pressures and temperatures discussing the stoichiometric effect ($\textnormal x < 1.0$) of hydrogen on the conductivity in the Earth’s core. The effect of increased hydrogen content results in lowering the resistivity for the \textit{fcc} phase. 
Yamakata~\etal~\cite{yamakata1994electrical} experimentally showed that resistivity increases (decrease in conductivity) due to hydrogen dissolution in iron with differing behavior in the temperature dependence compared to pure iron. However, there is a lack of experimental data and theoretical predictions for various phases at high pressures considering ambient temperature. 
Our study serves as an initial benchmark for future studies, notably, no experimental or theoretical data exist for direct comparison under these conditions. Note that a significant contribution of the total electrical conductivity at ambient temperature stems from the electronic component (80--90\%). While the contributions of the phononic and magnonic components are neglected here, they could be resolved~\cite{Nikolov2022}.

\begin{figure} 
\centering       
\includegraphics[width=1.0\columnwidth]{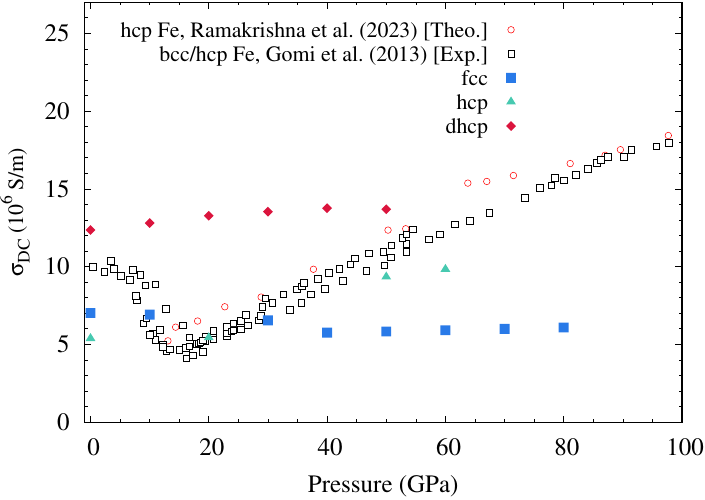}
    \caption{Pressure versus DC conductivity for \textit{hcp},  \textit{dhcp}, \textit{fcc} and N1 phases of FeH. Experimental data for \textit{bcc} and \textit{hcp} Fe stems from Ref.~\cite{gomi2013high}. Theoretical data for \textit{hcp} Fe stems from Ref.~\cite{Ramakrishna2023iop}.}
\label{fig:phase_eleccond}
\end{figure}

\cref{fig:phase_eleccond} shows the DC conductivity as a function of pressure for various FeH and Fe phases. 
As pressure increases, the conductivity drops in \textit{bcc} Fe with a further increase after a phase transformation to the \textit{hcp} phase around $\sim$13 GPa~\cite{gomi2013high,seagle2013electrical}. 
The theoretical results of \textit{hcp} iron as a comparison are indicated by empty red circles \cite{Ramakrishna2023iop} consistent with experimental DAC measurements reported for the \textit{hcp} Fe. 
Considering \textit{hcp} FeH, our results indicate the conductivity increases with pressure similar to \textit{hcp} Fe albeit with a lower magnitude across higher pressures due to the concentration of hydrogen in the system consistent with the lower electrical conductivity prediction in FeH by Yamakata~\etal~\cite{yamakata1994electrical}. 
Similarly, the conductivity increases with pressure for the \textit{dhcp} phase resulting in a larger DC conductivity due to the higher $c/a$ ratio ($\sigma_{\mathbf{zz}}$ component) compared to \textit{hcp} phase (see Figure S3 showing frequency-dependent electrical conductivity in the Supplemental Material~\cite{sup_mat}). 
The \textit{fcc} phase of FeH exhibits a decrease in conductivity with pressure similar to \textit{bcc} Fe albeit the change with pressure is smaller compared to \textit{hcp} FeH.
It is important to note that we did not compute conductivity for the novel FeH structures in this study due to the high computational costs involved.

\section{Conclusion and Outlook}
Using a machine learning approach, we have developed a highly transferable NN interatomic potential to explore the energy landscape of FeH across a broad pressure range from 0 to 100 GPa. By employing the combination of the MH method and a NN potential, we have extensively investigated the phase diagram of FeH, focusing on large-scale structure prediction, which is otherwise computationally highly demanding.

Throughout our search, we successfully discovered all the well-known bulk structures of FeH, demonstrating the reliability and accuracy of our approach. We also found many low-enthalpy structures at each pressure, which are modifications or stacking of the known \textit{dhcp}, \textit{hcp}, and \textit{fcc} phases. These new structures are energetically competitive with the most stable structures at each pressure. Our phonon calculations indicate that these structures are dynamically metastable and have a phonon DOS similar to \textit{dhcp} and \textit{hcp} at low frequencies. 
Our calculations show that the bulk modulus of the new phase N1 is very close to that of \textit{dhcp} and \textit{hcp} at low temperatures, but different at high temperatures. 
Also, we showed that the lattice parameters and EoS for the \textit{dhcp} phase agree with previous experimental results.
Furthermore, our calculations for the three known phases of FeH show that the electrical conductivity of the \textit{dhcp} and \textit{hcp} phases increases with pressure, in line with previous theoretical and experimental results for \textit{hcp} Fe. 
However, the conductivity of the \textit{fcc} phase decreases with pressure, similar to \textit{bcc} Fe.

The methodology employed in this study opens up possibilities to systematically investigate the phase diagram and PES of iron superhydrides \fh{n} ($n\ge3$), which exhibit remarkable electrical properties such as superconductivity. We expect that this computational approach will enable the exploration of novel and intriguing phenomena in these materials.

\begin{acknowledgments}
    We acknowledge fruitful discussions on \textit{ab-initio} molecular dynamics with Mandy     Bethkenhagen at the initial stages of this project. Also, H.T. thanks S. Alireza Ghasemi and Hossein Mirhosseini for valuable expert discussions in training the NN interatomic potentials.
    This work was partially supported by the Center for Advanced Systems Understanding (CASUS) which is financed by Germany’s Federal Ministry of Education and Research (BMBF) and by the Saxon state government out of the State budget approved by the Saxon State Parliament. 
    Computations were performed on a Bull Cluster at the Center for Information Services and     High-Performance Computing (ZIH) at Technische Universit\"at Dresden and on the cluster Hemera of the Helmholtz-Zentrum Dresden-Rossendorf (HZDR).
\end{acknowledgments}

%
%

\bibliography{main}

\end{document}